\documentclass[aps,pra,twocolumn,showpacs,preprintnumbers,superscriptaddress]{revtex4}
\usepackage{graphicx,bm,dcolumn,amsmath,amssymb,amsfonts}
\newcommand{\vm}{\bm{m}}

\newcommand{\vn}{\bm{n}}
\newcommand{\vsigma}{\bm{\sigma}}
\newcommand{\tr}{{\rm tr}}
\begin{document}

\title{Composite pulses in NMR as non-adiabatic geometric quantum gates} 

\author{Yukihiro Ota}
\altaffiliation[Present address:]{
CCSE, Japan Atomic Energy Agency, 
6-9-3 Higashi-Ueno, Tokyo 110-0015, Japan 
and 
CREST(JST), 
4-1-8 Honcho, Kawaguchi, Saitama, 332-0012, Japan}
\affiliation{
Research Center for Quantum Computing, Interdisciplinary Graduate School 
of Science and Engineering, Kinki University, 
3-4-1 Kowakae, Higashi-Osaka, 577-8502, Japan}
\author{Yasushi Kondo}
\affiliation{
Research Center for Quantum Computing, Interdisciplinary Graduate School 
of Science and Engineering, Kinki University, 
3-4-1 Kowakae, Higashi-Osaka, 577-8502, Japan}
\affiliation{
Department of Physics, Kinki University, 
3-4-1 Kowakae, Higashi-Osaka, 577-8502, Japan}

\date{\today}

\begin{abstract}
We show that some composite pulses widely employed in nuclear magnetic
 resonance experiments are
 regarded as non-adiabatic geometric quantum gates with Aharanov-Anandan
 phases. 
Thus, we reveal the presence of a fundamental issue on quantum mechanics
 behind a traditional technique.
To examine the robustness of such composite pulses against fluctuations,
 we present a simple noise model in a two-level system. 
Then, we find that the composite pulses possesses purely geometrical
 nature even under a certain type of fluctuations.  
\end{abstract}

\pacs{03.65.Vf, 82.56.-b, 82.56.Jn, 03.67.-a, 03.65.-w}
\maketitle

Geometric phases have been attracting a lot of attention from 
the view point of the foundation of quantum mechanics and mathematical
physics~\cite{Nakahara2003,ChruscinskiJamiolkowski2004,Vojta2003,BengtssonZyczkowski2006}. 
Recently, their application to quantum information processing is 
spotlighted~\cite{others,Tianetal2004}, because they
are expected to be robust against noise. 
However, the robustness of a geometric quantum gate (GQG), which is a
quantum gate only using geometric phases, is not completely
verified. 
Various examinations on this issue have been
reported~\cite{BlaisTremblay2003,ZhuZanardi2005,NazirSpillerMunro2002,Carollo;Vedral:2003,Chiara;Palma:2003,Dajka;Luczka:2008}.  
Blais and Tremblay~\cite{BlaisTremblay2003} claimed
that no advantage of the GQGs exists compared to the 
corresponding quantum gates with dynamical phases, while
Zhu and Zanardi~\cite{ZhuZanardi2005}
showed that their non-adiabatic GQGs are robust against 
fluctuations in control parameters. 

In this paper, we show that some composite pulses widely
employed in nuclear magnetic resonance (NMR)~\cite{Levitt1986,Claridge1999} to accomplish reliable
operations is regarded as non-adiabatic GQGs based on
an Aharonov-Anandan (AA) phase~\cite{AharonovAnandan1987}, and propose a
simple noise model in a two-level system. 
Then, we classify fluctuations in terms of the robustness of the GQGs. 

An AA phase appears under non-adiabatic cyclic time evolution of a
quantum system~\cite{AharonovAnandan1987}. 
We note that the generalization to the non-cyclic case is given in
Ref.~\cite{ChruscinskiJamiolkowski2004,noncg}. 
Let us write the Bloch vector at $t$ ($0\le t\le 1$) as 
$\vn(t)(\in\mathbb{R}^{3})$. 
We denote a state vector given $\vn(t)$ as
$|\vn(t)\rangle(\in\mathbb{C}^{2})$. 
Namely, 
\(
\vn(t) = \langle \vn(t)|\vsigma|\vn(t)\rangle
\), where 
$\vsigma=\!^{t}(\sigma_{x},\,\sigma_{y},\,\sigma_{z})$. 
The symbol $^{t}$ means the transposition of a vector. 
Time evolution is described by the Schr\"odinger equation
with the Hamiltonian $H(t)$.   
Note that $|\vn(t)| = 1$.  
Hereafter, we denote $\vn(0)$ as $\vn$.  
We take the natural unit system in which $\hbar=1$. 
Suppose that \(|\vn(1)\rangle = e^{i\gamma}|\vn\rangle\)
($\gamma\in\mathbb{R}$): 
\(
\vn(1)=\vn
\). 
The AA phase $\gamma_{{\rm g}}$ is defined as\,\cite{AharonovAnandan1987}
\begin{equation}
\gamma_{{\rm g}} = \gamma - \gamma_{{\rm d}},
\label{eq:def_AAphase}
\end{equation}
where 
\begin{equation}
\gamma_{{\rm d}} 
= -\int^{1}_{0}\langle \vn(t)|H(t)|\vn(t)\rangle\, dt
\label{eq:dyn_phase}
\end{equation}
is a dynamical phase. 

Next, suppose $\vn_{+}$ and $\vn_{-}$ are two Bloch vectors satisfying (a)
$\vn_{+}\cdot\vn_{-}=-1$ (i.e., $\langle \vn_{+}|\vn_{-}\rangle =0$) and
(b) $\vn_{\pm}(1) = \vn_{\pm}$ (i.e., there exist
$\gamma_{\pm}\in\mathbb{R}$ such that 
\(
|\vn_{\pm}(1)\rangle = e^{i
\gamma_{\pm}} |\vn_{\pm} \rangle 
\). 
An arbitrary quantum state $|\vn \rangle$ is expressed by 
\(
|\vn \rangle = a_{+}|\vn_{+}\rangle + a_{-}|\vn_{-}\rangle
\), where 
\(
a_{\pm}=\langle \vn_{\pm}|\vn\rangle
\). 
We call $\vn_{\pm}$ 
basis Bloch vector corresponding to $H(t)$.
The initial state $|\vn\rangle$ is transformed into the final state 
\(
|\vn(1)\rangle  
=  a_{+}\,e^{i \gamma_{+}}|\vn_{+}\rangle 
+  a_{-}\,e^{i \gamma_{-}}|\vn_{-}\rangle
\). 
Thus, the time evolution operator $U$ at $t=1$ generated by $H(t)$
($t\in [0,1]$) is rewritten as
\begin{equation}
 U
= 
 e^{i\gamma_{+}}|\vn_{+}\rangle\langle \vn_{+}|
+
 e^{i\gamma_{-}}|\vn_{-}\rangle\langle \vn_{-}|.
\label{eq:geometricgate}
\end{equation}
Equation (\ref{eq:geometricgate}) becomes a quantum gate with a
geometric phase, when the dynamical component of $\gamma_{\pm}$ is
vanishing. 

Let us focus on the Hamiltonian for a one-qubit system,  
\begin{equation}
 H(t) = \frac{1}{2}\omega(t)\, \vm(t) \cdot \vsigma
\qquad  (0 \le t \le 1),
\label{eq:Hamiltonian}
\end{equation}
which is inspired by a NMR Hamiltonian. 
In the case of NMR, $\omega(t)$ and $\vm(t)$ are the amplitude of and
the unit vector parallel to a magnetic field, respectively.   
The dynamical phase vanishes when 
$\vm(t) \cdot \vn(t)=0$~\cite{SuterMuellerPines1988}. 
We note that the integrand in Eq.\,(\ref{eq:dyn_phase}) is
rewritten as 
\(
\langle \vn(t) | H(t) |\vn(t)\rangle
=
(\omega(t)/4)\tr[(\vm(t)\cdot\vsigma)(\vn(t)\cdot\vsigma)]
=
(\omega(t)/2)\vm(t)\cdot\vn(t)
\), 
where we use $\tr[H(t)]=0$ and 
$\tr (\sigma_{i}\sigma_{j})=2\delta_{ij}$. 
This condition has been widely used in the experiments 
on non-adiabatic GQGs~\cite{Tianetal2004}. 

A series of pulses, $90_{x}180_{y}90_{x}$ has been widely employed in the 
field of NMR for wide band decoupling~\cite{Levitt1986,Claridge1999}, 
where $\beta_{k}$ denotes a spin rotation by the angle 
$\beta$ in degree around $k$-axis. 
This is called composite pulse and corresponds to the unitary operator  
\(
e^{-i\pi\sigma_{x}/4}e^{-i\pi\sigma_{y}/2}e^{-i\pi\sigma_{x}/4}
\), which is equal to $e^{-i\pi\sigma_{y}/2}$. 
This is generated by the Hamiltonian 
\begin{equation}
H(t)
= \pi \vm(t) \cdot \vsigma
\quad
(0\le t\le 1),
\label{mlev_sq}
\end{equation} 
where
\begin{equation*}
 \vm(t)
=\left\{
\begin{array}{ccl}
\!^{t}(1,0,0) & & (0\le t \le 1/4) \\
\!^{t}(0,1,0) & & (1/4\le t\le 3/4)\\
\!^{t}(1,0,0) & & (3/4\le t\le 1)
\end{array}
\right. .
\end{equation*}
Hereafter, we will denote 
$t_{0}=0$, $t_{1}=1/4$, $t_{2}=3/4$, and
$t_{3}=1$. 
Various types of composite pulses have been
proposed\,\cite{Levitt1986,Claridge1999}, and their usages 
have been also discussed in the context of NMR quantum
computing\,\cite{cp_qip}. 

\begin{figure}[tp]
(a)\!\scalebox{0.72}[0.72]{\includegraphics{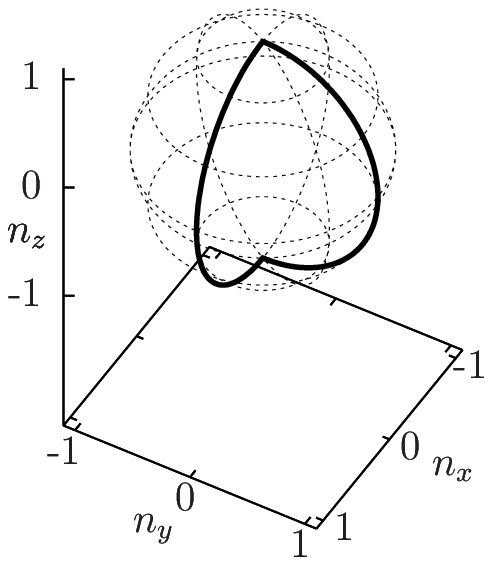}}
(b)\!\scalebox{0.72}[0.72]{\includegraphics{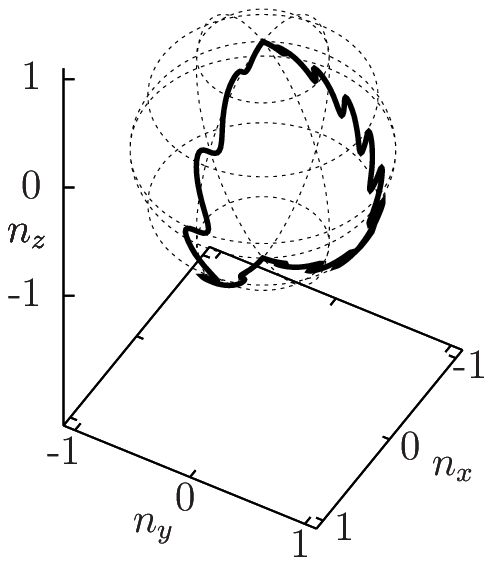}}
\caption{\label{fig:mlev}
Temporal behavior of the 
basis Bloch vector $\!^{t}(0,\,1,\,0)$
during the composite pulse $90_{x}180_{y}90_{x}$. 
(a) without 
and (b) with fluctuations in the control parameters. 
The fluctuations are given by Eq.~(\ref{eq:noise_func_num}), 
where $f_{0}=g_{0}=0.1$ and $\xi=\eta=5$. }
\end{figure}
Let us examine the time evolution generated by Hamiltonian
(\ref{mlev_sq}) from the view point of non-adiabatic GQGs. 
We choose $\vn_\pm=\!^{t}(0,\pm 1,0)$, where $\vn_{+}\cdot\vn_{-}=-1$. 
Then, we have the explicit formula
\begin{eqnarray}
\vn_{\pm}(t) 
=
\pm
\left(
\begin{array}{c}
 \sin \theta(t)\sin \phi(t) \\
-\sin \theta(t)\cos \phi(t) \\
 \cos \theta(t)
\end{array}
\right),
\label{eq:path_mlev} 
\end{eqnarray}
where 
\begin{equation*}
 \theta(t) = 2\pi t - \frac{\pi}{2},
\quad
 \phi(t)
=\left\{
\begin{array}{ccl}
\pi/2 & & (t_{1}\le t\le t_{2}) \\
0 & & (\text{otherwise})
\end{array}
\right. .
\end{equation*}
The temporal behavior of $\vn_{+}$ on the Bloch sphere is shown in
Fig.\,\ref{fig:mlev}(a). 
The trajectory $\vn_{+}$ is closed. 
It means that 
$|\vn_{+}(1)\rangle = e^{i\gamma_+}|\vn_{+}\rangle$. 
We find that 
\(
|\vn_\pm(1)\rangle=e^{\mp i \pi /2} |\vn_{\pm} \rangle
\) via solving the Schr\"odinger equation. 
We note that $\pi$ is a solid angle surrounded by the trajectory
$\vn_{+}(t)$. 
We also find that $\vm(t)\cdot\vn_{\pm}(t)=0$ at any $t\in[0,\,1]$, and
thus the dynamical component is vanishing.  
Accordingly, we obtain the non-adiabatic GQG, 
\(
U
=  e^{-i \pi/2} |\vn_{+}\rangle \langle \vn_{+}| 
 + e^{ i \pi/2} |\vn_{-}\rangle \langle \vn_{-}| 
= e^{-i \pi \sigma_{y}/2}
\). 
One of the most commonly employed composite pulses turns out a
non-adiabatic GQG\,\cite{waltz}.

We will classify fluctuations in terms of robustness of 
the composite pulse $90_{x}180_{y}90_{x}$. 
A noise model will be proposed based on a fluctuated closed curve on the
Bloch sphere. 
We examine the situation in which the radio-frequency (rf) amplitude and phase, and the
resonance off-set are temporary fluctuated around their aimed values. 
The fluctuated curve is given by 
\begin{equation}
\tilde{\vn}_{\pm}(t)
=
\pm
\left(
\begin{array}{c}
 \sin(\theta(t)+ f(t))\,\sin(\phi(t) + g(t)) \\
-\sin(\theta(t)+ f(t))\,\cos(\phi(t) + g(t)) \\
 \cos(\theta(t)+ f(t))
\end{array}
\right),
\label{eq:noise_Bloch} 
\end{equation}
where we assume that $f(t)$ and $g(t)$ are continuous and smooth in
$[0,\,1]$~\cite{remark_pw_cont} and satisfy  
\begin{equation}
 f(t_{0})=g(t_{0})=0,\quad
 f(t_{3})=g(t_{3})=0. 
\label{eq:noise_endpoint}
\end{equation}
We will discuss the relevance of $f(t)$ and $g(t)$ to 
fluctuations below. 
The trajectory $\tilde{\vn}_{\pm}(t)$ is closed under the assumption
(\ref{eq:noise_endpoint}), as shown in Fig.\,\ref{fig:mlev}(b). 
Tnus, we have 
\begin{equation}
 |\tilde{\vn}_{\pm}(1)\rangle 
= e^{i\tilde{\gamma}_{\pm}}|\tilde{\vn}_{\pm}\rangle,
\end{equation}
with a phase $\tilde{\gamma}_{\pm}$.  
Generally, $\tilde{\gamma}_{\pm}$ includes both the dynamical
and the geometric components. 
We employ this noise model in order to ensure 
the existence of a definite AA phase, although we aware of its artificiality. 
An analysis based on a non-cyclic geometric
phase\,\cite{Dajka;Luczka:2008,noncg} may be needed for more
comprehensive discussions. 

We derive the Hamiltonian generating the time evolution corresponding to
Eq.~(\ref{eq:noise_Bloch}). 
By differentiating Eq.~(\ref{eq:noise_Bloch}) with respect to  
$t \in (t_{i-1},t_{i})$ ($i=1,\,2,\,3$), we obtain the Bloch equation.
Then, we find the Hamiltonian in this time interval.
Hence, the Hamiltonian at $t\in [0,1]$ is given by
\begin{eqnarray}
\tilde{H}(t) 
&=& \frac{1}{2}
\tilde{\omega}(t)\, \tilde{\vm}(t)\cdot\vsigma  
+  \frac{1}{2}\frac{dg(t)}{dt} \sigma_{z},
\label{eq:noiseHamiltonian}
\end{eqnarray}
where
\begin{equation*}
\tilde{\omega}(t)
=
2\pi + \frac{df(t)}{dt},\quad
\tilde{\vm}(t)
=
\left(
\begin{array}{c}
\cos(\phi(t) + g(t)) \\
\sin(\phi(t) + g(t)) \\
0
\end{array}
\right). 
\end{equation*}
We find that 
\begin{equation}
 \tilde{\vm}(t)\cdot \tilde{\vn}(t) =0. 
\label{eq:inn_p_noise}
\end{equation}
at any $t \in [0,1]$. 
The derivative of $f(t)$ is a fluctuation of 
the rf amplitude, while that of $g(t)$ is 
that of the resonance off-set. 
A fluctuation of the rf phase is described by $g(t)$. 
From Eq.\,(\ref{eq:dyn_phase}), the dynamical component
$\tilde{\gamma}_{{\rm d}\pm}$ of $\tilde{\gamma}_\pm$ is given by 
\begin{eqnarray}
 \tilde{\gamma}_{{\rm d}\pm} 
&=&
\mp \frac{1}{2}
\int_{t_{0}}^{t_{3}}\frac{dg(t)}{dt}
\cos[\theta(t)+ f(t)]
\,dt.
\label{eq:explicite_dyn_phase_noise}
\end{eqnarray}

\begin{figure}[tp]
\centering
(a)\!\!\!\scalebox{0.47}[0.47]{\includegraphics{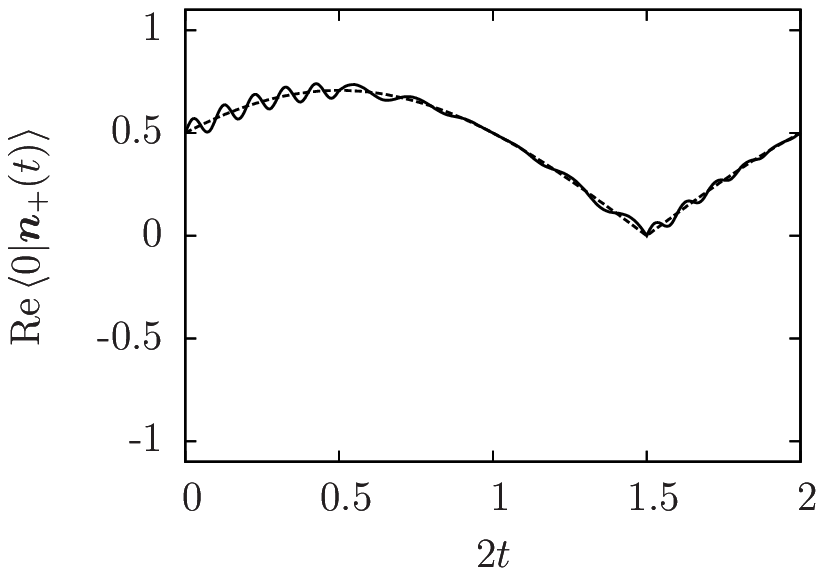}}
(b)\!\!\!\scalebox{0.47}[0.47]{\includegraphics{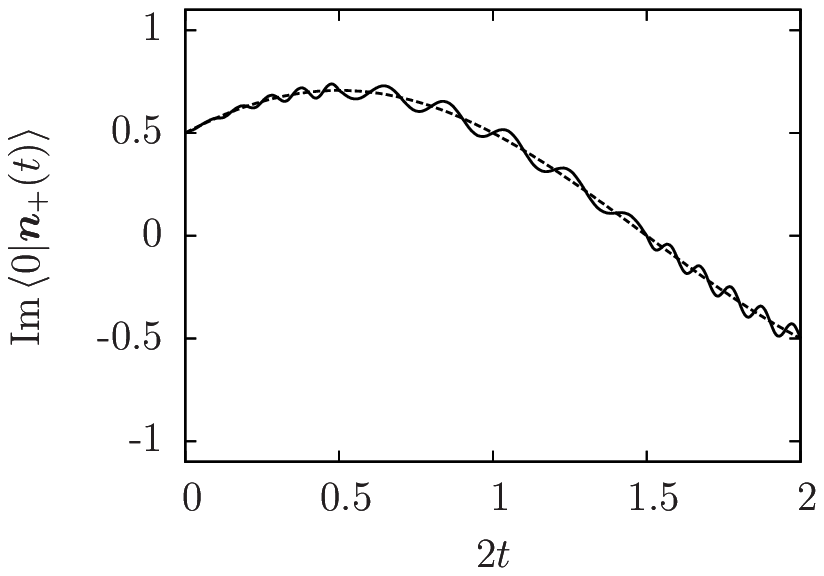}}
(c)\!\!\!\scalebox{0.47}[0.47]{\includegraphics{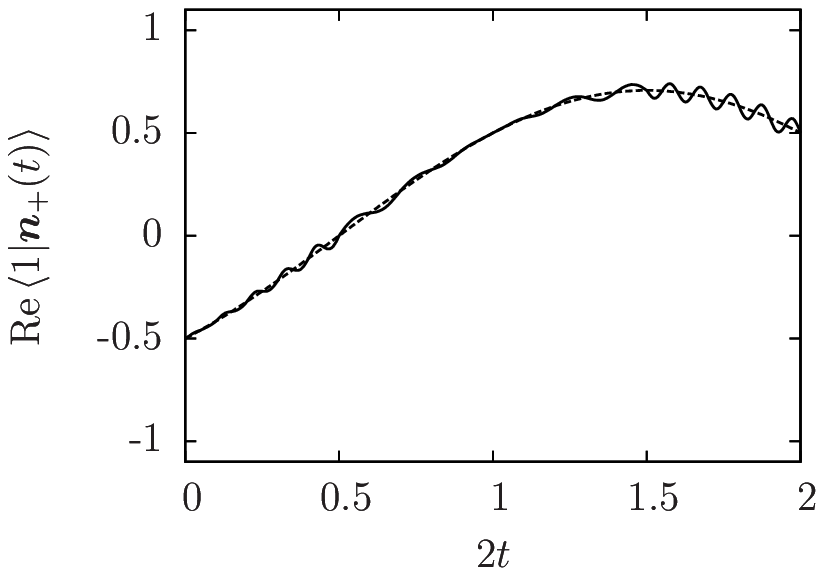}}
(d)\!\!\!\scalebox{0.47}[0.47]{\includegraphics{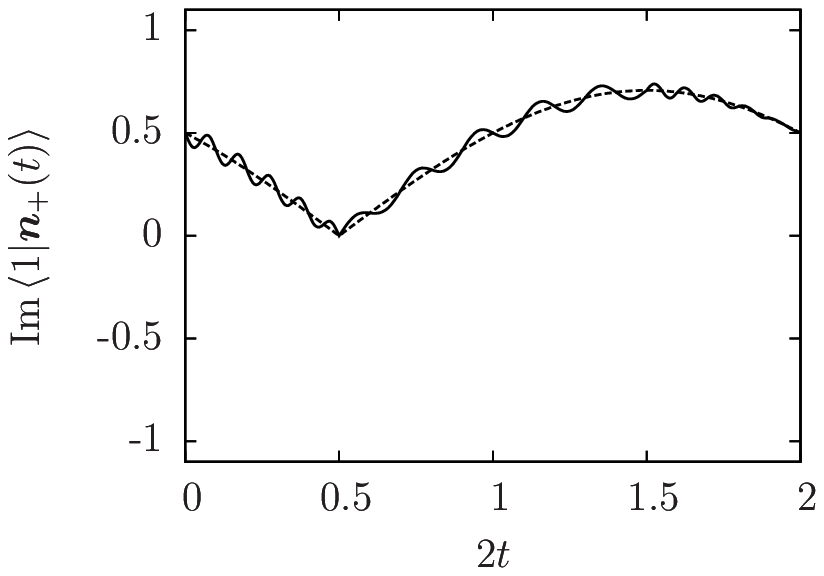}}
\caption{\label{fig:mlev_amp} 
Temporal behavior of the state vector corresponding
to the basis Bloch vector $\,^{t}(0,\,1,\,0)$ during 
$90_{x}180_{y}90_{x}$. 
The initial state vectors are chosen as 
\(
|\vn_{+}\rangle = e^{i\pi/4}(|0\rangle + i|1\rangle)/\sqrt{2}
\).
The solid line is the model with the fluctuations.  
The fluctuations are described by Eq.~(\ref{eq:noise_func_num}), 
where $f_{0}=g_{0}=0.1$ and $\xi=\eta=5$. 
The dashed line is the ideal case. 
(a) ${\rm Re}\langle 0|\vn_{+}(t)\rangle$.
(b) ${\rm Im}\langle 0|\vn_{+}(t)\rangle$. 
(c) ${\rm Re}\langle 1|\vn_{+}(t)\rangle$.
(d) ${\rm Im}\langle 1|\vn_{+}(t)\rangle$. }
\end{figure}

We show that the following two cases exactly lead to 
\(
\tilde{\gamma}_{{\rm d}\pm} =0
\). 
Namely, (i) $g(t)=0$ and (ii) $f(t)$ and $g(t)$ have a 
certain symmetric property under time translation. 
The validity of the case (i) is obvious from
Eq.~(\ref{eq:explicite_dyn_phase_noise}). 
We focus on the case (ii). 
We note that $90_{x}180_{y}90_{x}$ has
several interesting properties under time translation: 
\(
\theta(t+1/2) = \theta(t) +\pi
\), for example. 
We divide the total time interval 
$I_{\rm all}=\{t\in [t_{0},t_{3}]\}$ into the four intervals, 
\(
I_{1}=\{t\in [t_{0},t_{1}]\}
\), 
\(
I_{2}=\{t\in [t_{1},1/2]\}
\), 
\(
I_{3}=\{t\in [1/2,t_{2}]\}
\), and 
\(
I_{4}=\{t\in [t_{2},t_{3}]\}
\). 
Let us consider a case when the conditions 
\begin{equation}
 f(t+1/2) = f(t),\quad
 \frac{dg}{dt}(t+1/2)=\frac{dg}{dt}(t),
\label{eq:sym_ts_one}
\end{equation}
are satisfied. 
The contribution from $I_{1}$ ($I_{2}$) to $\tilde{\gamma}_{{\rm d}\pm}$
is canceled out by that from $I_{3}$ ($I_{4}$). 
Thus, this case leads to $\tilde{\gamma}_{{\rm d}\pm}=0$. 
Let us consider another case, in which the conditions 
\begin{equation}
 f(1-t) = -f(t),\quad
 \frac{dg}{dt}(1-t)=\frac{dg}{dt}(t), 
\label{eq:sym_ts_two}
\end{equation}
are satisfied. 
We note that $f(1/2)=0$ is imposed in Eq.\,(\ref{eq:sym_ts_two}). 
In this case, the contribution from $I_{1}$ ($I_{2}$) is canceled out by
$I_{4}$ ($I_{3}$).   
This cancellation is related to the symmetry
$\theta(1-t)=-\theta(t)+\pi$.  
When $f(t)$ and $g(t)$ have a certain symmetric
property compatible with the pulse sequence, the dynamical phase is vanishing.
In addition, a case (iii) $f(t)$ and $g(t)$ rapidly oscillate with no
correlation, leads to $\tilde{\gamma}_{{\rm d}\pm} \approx 0$. 
We can confirm the validity of the case (iii) by numerically solving the
Schr\"odinger equation with Eq.\,(\ref{eq:noiseHamiltonian}). 
The case (i) often happens in experiments. 
From Eq.~(\ref{eq:noiseHamiltonian}), one can find $f(t)$ is associated
only with the amplitude of an external controlled field. 
This quantity often shows an overshoot or an undershoot before
settling a desired strength. 
One can also encounter the case (ii) in experiments. 
A typical example for Eq.\,(\ref{eq:sym_ts_one}) may be an oscillating
function, as shown in Eq.\,(\ref{eq:periodic_noise_example}). 
A linear combination of such oscillating functions leads to
$\tilde{\gamma}_{{\rm d}\pm}=0$. 
Thus, we expect that a lot of rapid oscillating fluctuations
approximately satisfy Eqs.(\ref{eq:sym_ts_one}) or
(\ref{eq:sym_ts_two}), and then 
\(\tilde{\gamma}_{{\rm d}\pm} \approx 0\).   
The case (iii) is natural when the origins of $f(t)$
and $g(t)$ are independent. 
These three conditions lead to  
\(
\tilde{\gamma}_{{\rm d}\pm}=0
\). 
Thus, the quantum gate under them is still regarded as a GQG.  
It is necessary to examine about more realistic 
control processes\,\cite{Mehring;Waugh:1972,Ryan;Laflamme:2009}. 
Nevertheless, the present discussion is meaningful 
to understand nature of robustness of a geometric phase. 

We directly solve the Schr\"odinger equation with
Eq.\,(\ref{eq:noiseHamiltonian}) in order to calculate the geometric
component of $\tilde{\gamma}_{\pm}$. 
First, we choose
\begin{eqnarray}
f(t)= f_{0}\sin [2\pi\xi u_{i}(t)],\quad
g(t)= g_{0}\sin [2\pi\eta u_{i}(t)],
 \label{eq:noise_func_num}
\end{eqnarray} 
at $t\in[t_{i-1},\,t_{i}]$, where 
$u_{i}(t)=(t-t_{i-1})/(t_{i}-t_{i-1})$ and $\xi,\,\eta \in\mathbb{N}$. 
The above functions are piecewise smooth in
$[t_{0},t_{3}]$\,\cite{remark_pw_cont}. 
We show that the temporal evolution of the basis Bloch vector 
$^{t}(0,1,0)$ during the composite pulse $90_{x}180_{y}90_{x}$ 
with the fluctuations in Fig.\,\ref{fig:mlev}(b). 
This example corresponds to the case (ii), since 
Eq.\,(\ref{eq:sym_ts_two}) is satisfied. 
We display the temporal behaviors of $|\vn_{+}(t)\rangle$ and
$|\tilde{\vn}_{+}(t)\rangle$ in Fig.\,\ref{fig:mlev_amp}.  
The state vector $|\tilde{\vn}_{+}(t)\rangle$ is fluctuated 
around $|\vn_{+}(t)\rangle$, but
$|\tilde{\vn}_{+}(t_3)\rangle = |\vn_{+}(t_3)\rangle$.
We find that $\tilde{\gamma}_{\pm}=\mp \pi/2$.
Thus, $\tilde{\gamma}_{{\rm g}\pm}=\mp \pi/2$ is confirmed. 
Let us discuss another example,
\begin{equation}
 f(t)=f_{0}\sin (8\pi\xi t),\quad
 g(t)=g_{0}\sin (8\pi\eta t),
\label{eq:periodic_noise_example}
\end{equation}
where  $f_{0}$ ($g_{0}$) is a positive real number and $\xi$
($\eta$) is an integer ($t_0\le t\le t_3$). 
The above functions also satisfy Eq.~(\ref{eq:noise_endpoint}). 
Solving the Schr\"odinger equation numerically leads to 
$\tilde{\gamma}_{\pm}=\tilde{\gamma}_{{\rm g}\,\pm}=\mp\pi/2$.
The above results mean that the solid angle surrounded by $\tilde{\vn}_{\pm}(t)$
is always $\pi$. 
We conjecture that, as long as
the fluctuations are introduced by Eqs.~(\ref{eq:noise_Bloch}) and
(\ref{eq:noise_endpoint}), no dynamical phase should exactly lead to 
\(
\tilde{\gamma}_{{\rm g}\pm} =
\gamma_{{\rm g}\pm}
\). 

It is interesting to study the case in which $\vm(t) \cdot \vn(t) \ne 0$.  
Let us consider a simple operation 
on the Bloch sphere: 
\(
\!^{t}(0,\,0,\,1)\, \to \!^{t}(1,\,0,\,0)
\). 
This process is realized by using either 
\(
e^{-iH_{{\rm A}}t}
\) or 
\(
e^{-iH_{{\rm B}}t}
\) ($0\le t\le 1$), where 
\(
H_{{\rm A}} = \pi\sigma_{y}/4
\) and 
\(
H_{{\rm B}} = \pi(\sigma_{x}+\sigma_{z})/2\sqrt{2}
\). 
The former satisfies the condition 
$\vm(t)\cdot\vn(t)=0$, but the latter does not.  
We describe fluctuations in the two models such as Eq.\,(\ref{eq:noiseHamiltonian}),   
\begin{eqnarray*}
 \tilde{H}_{{\rm A}}(t)
&=&
\left(\frac{\pi}{2}+\frac{df}{dt}\right)
\frac{\tilde{\vm}_{{\rm A}}(t)\cdot\vsigma }{2}
+ \frac{dg}{dt}\frac{\sigma_{z}}{2}, \\
 \tilde{H}_{{\rm B}}(t)
&=&
\left(\frac{\pi}{\sqrt{2}}+\frac{df}{dt}\right)
\frac{\tilde{\vm}_{{\rm B}}(t)\cdot\vsigma }{2} 
+ \left(\frac{\pi}{\sqrt{2}}+\frac{dg}{dt}\right)\frac{\sigma_{z}}{2}, 
\end{eqnarray*} 
where 
\(
 \tilde{\vm}_{{\rm A}}(t)
=
\!^{t}\left(
\cos(\pi/2+g(t)), 
\sin(\pi/2+g(t)), 
0
\right)
\) and 
\(
 \tilde{\vm}_{{\rm B}}(t)
=
\!^{t}\left(
\cos g(t), 
\sin g(t), 
0
\right)
\). 
\begin{figure}[tp]
(a)\!\scalebox{0.61}[0.61]{\includegraphics{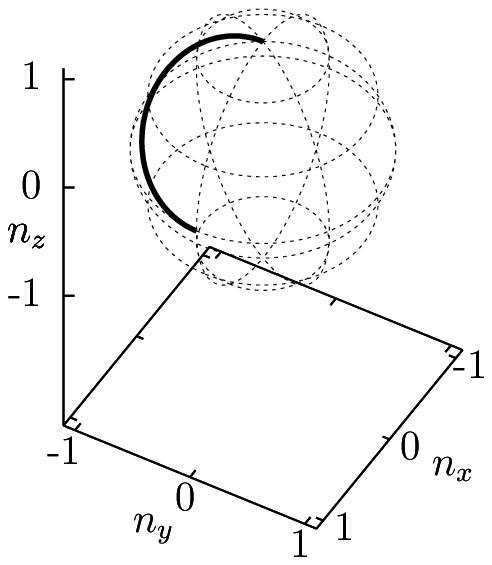}}
(b)\!\scalebox{0.61}[0.61]{\includegraphics{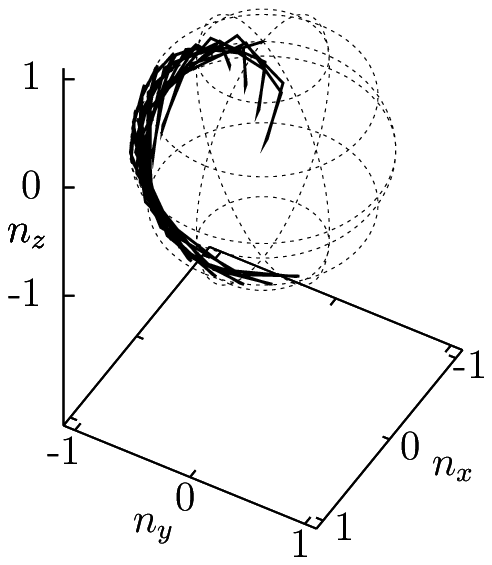}}\\\vspace{5mm}
(c)\!\!\!\scalebox{0.53}[0.53]{\includegraphics{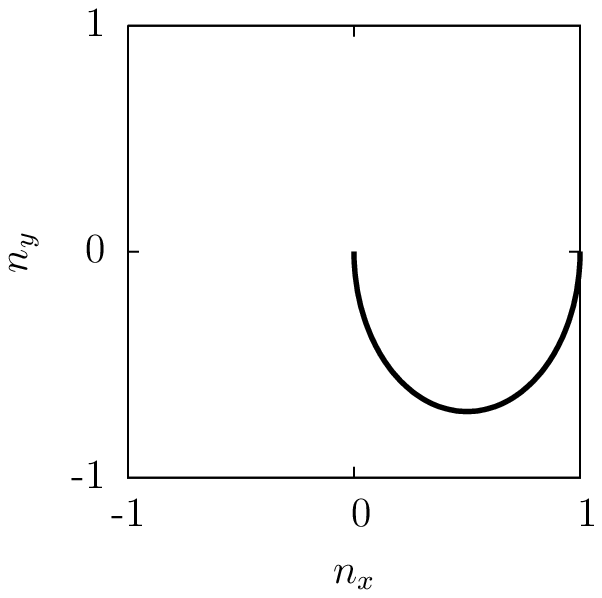}}
(d)\!\!\!\scalebox{0.53}[0.53]{\includegraphics{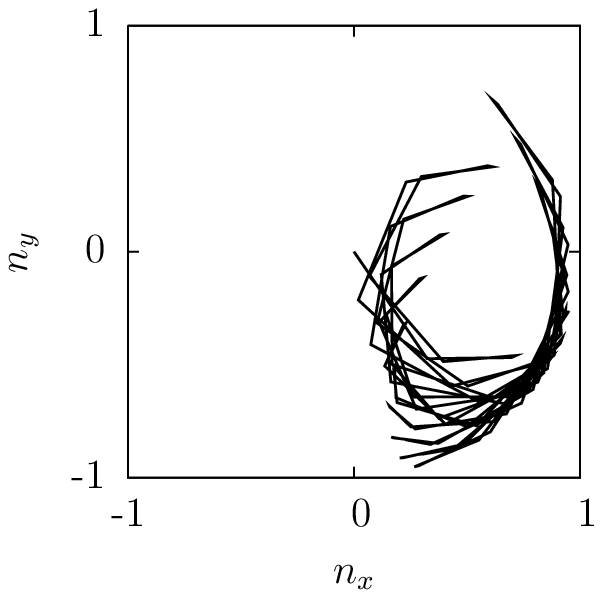}}
\caption{\label{fig:another_model}
Temporal behavior of the Bloch vector starting from $\!^{t}(0,0,1)$
under the Hamiltonian $H_{\rm B}$ is shown in (a) and its trajectory
projected on $n_x n_y$-plane is shown in (c).
The final point is $\!^{t}(1,0,0)$. 
Temporal behavior of the Bloch vector starting from $\!^{t}(0,0,1)$ under
the fluctuating Hamiltonian $\tilde{H}_{\rm B}$
($f_0 = g_0 = 1.0$ and $\xi = \eta = 10$ in
 Eq.\,(\ref{eq:periodic_noise_example})) 
is show in (b) and its trajectory projected on $n_x n_y$-plane
is shown in (d). The final point is $\!^{t}(0.95, -0.26, -0.16)$. }
\end{figure}
Since $f(0) = f(1) = g(0) =g(1) =0$, which corresponds to
Eq.\,(\ref{eq:noise_endpoint}), the unitary operator generated by 
$\tilde{H}_{\rm A}(t)$ maps $\!^{t}(0,0,1) \to \!^{t}(1,0,0)$ even in the
presence of $f(t)$ and $g(t)$. 
On the other hand, the numerical calculation reveals that the one
generated by $\tilde{H}_{\rm B}(t)$ maps 
$\!^{t}(0,0,1) \to \!^{t}(0.95,-0.26,-0.16)$ [Fig.\,\ref{fig:another_model}].   
The results mean that Eq.\,(\ref{eq:noise_endpoint}) does
not always ensure robustness in the present model. 
We can find an additional term appears in Eq.\,(\ref{eq:path_mlev}) when
$\vm(t)\cdot \vn(t) \neq 0$. 
Thus, it may cause a strong fluctuation. 
We guess that $\vm(t)\cdot \vn(t)=0$ might play an 
important role for stable time evolution in the present model. 

In conclusion, we showed that the composite pulse $90_{x}180_{y}90_{x}$
is regarded as a non-adiabatic GQG.  
In addition, we proposed a simple noise model based on a fluctuated
curve on the Bloch sphere, and then 
classified fluctuations in terms of robustness of $90_{x}180_{y}90_{x}$.
Although the present analysis is artificial, it is
suitable for evaluating errors in non-adiabatic GQGs since
a definite geometric phase exists even in the presence of fluctuations. 
It is important to improve the present method in order to examine a
more realistic control process or a stochastic process. 
The fluctuations that we discussed 
should be called regular fluctuations, because the fluctuations 
are expressed by the two smooth functions $f(t)$ and $g(t)$. 
On the other hand, when fluctuations are given by uniform random
variables, even a cyclic evolution may not be guaranteed~\cite{supercycle} 
and thus the robustness is not expected as discussed in
Ref.\,\cite{BlaisTremblay2003}. 
We emphasize that it is important to specify fluctuations in order
to evaluate robustness of a gate. 

The authors wish to acknowledge helpful discussion with M. Nakahara. 
This work was supported by ``Open Research Center'' Project for
 Private Universities: Matching fund subsidy from Ministry of
 Education, Culture, Sports, Science and Technology.


\begin{thebibliography}{99}
\bibitem{remark_pw_cont}
When $f(t)$ and $g(t)$ are piecewise smooth, the current
	discussion is also possible. 
We should divide the time interval into smaller pieces, in
	which they are smooth.
\end{thebibliography}
\end{document}